# MP2-F12 Basis Set Convergence for the S66 Noncovalent Interactions Benchmark: Transferability of the Complementary Auxiliary Basis Set (CABS)

Nitai Sylvetsky[1], Manoj Kumar Kesharwani,[1] and Jan M.L. Martin[1, a)]

[1] *Department of Organic Chemistry, Weizmann Institute of Science, 76100 Reḥovot, Israel*

a) Corresponding author: gershom@weizmann.ac.il

**Abstract.** Complementary auxiliary basis sets for F12 explicitly correlated calculations appear to be more transferable between orbital basis sets than has been generally assumed. We also find that aVnZ-F12 basis sets, originally developed with anionic systems in mind, appear to be superior for noncovalent interactions as well, and propose a suitable CABS sequence for them.

## INTRODUCTION

Explicitly correlated methods (see[1,2] for recent reviews) are extremely powerful tools for thermochemistry (e.g.,[3,4]), computational spectroscopy (e.g.,[5]), and the study of noncovalent interactions.[6] The latter will be the focus of the present paper.

Explicitly correlated MP2-F12 or CCSD(F12*) calculations in practical implementations require not only the specification of an orbital basis set and a geminal exponent, but also of three auxiliary basis sets (which we will denote by their acronyms in the MOLPRO program system[7]) that allow avoiding the need for three-and four-electron integrals:

• a "JKfit" basis set for the expansion of Coulomb (J) and exchange (K) type integrals, such as also used in density fitting Hartree-Fock or hybrid DFT calculations;

• an "MP2fit" basis set for the RI-MP2 approximation, such as also used in density fitting MP2 (a.k.a. resolution of the identity MP2, RI-MP2) or double-hybrid DFT calculations;

• an "OptRI" or "CABS" (complementary auxiliary basis set[8]) used for the evaluation of the F12-specific matrix elements.

For the most widely used orbital basis set families, there are optimized Standard JKfit[9] and MP2fit[10,11] basis sets available, while for the two most commonly used families employed inside explicitly correlated calculations, namely augmented correlation consistent aug-cc-pVnZ (n=D,T,Q,5;[12,13] AVnZ for short) and F12-correlation consistent cc-pVnZ-F12 (n=D,T,Q),[14] compact aug-cc-pVnZ/OptRI and cc-pVnZ-F12/OptRI CABS basis sets, respectively,[15,16] have been optimized. Very recently, Shaw and Hill published[17] improved cc-pVnZ-F12/OptRI+ basis sets that add a few extra functions optimized for the CABS-corrected HF energy. For the largest F12-optimized orbital basis set, cc-pV5Z-F12,[18] no OptRI is as yet available: the original paper[18] recommends using the aug-cc-pwCV5Z/MP2fit basis set instead as a brute-force alternative.

It was during a recent re-examination of the S66 benchmark[19,20] for noncovalent interactions that we considered the ano-pVnZ+ basis sets of Valeev and Neese[21] as an alternative. For these, no optimized CABS basis sets are available, which led us to wonder about the transferability of OptRI basis sets between families. A preliminary report of our findings is presented here.

## COMPUTATIONAL METHODS

Most calculations were performed using the MOLPRO 2015.1 program system[7] running on the Faculty of Chemistry HPC facility.

The cc-pVnZ-F12 (VnZ-F12 for short) basis sets only carry diffuse s and p functions. For applications on anionic systems, we recently developed[24] the aug-cc-pVnZ-F12 (aVnZ-F12 for short) basis sets, which have added diffuse functions of d and higher angular momenta as well. We will also consider their performance here.

## RESULTS AND DISCUSSION

### MP2-F12 Basis Set Convergence

First, let us consider the valence double-zeta (VDZ) basis set. sano-pVDZ is clearly unacceptably small, while ordinary aug-pVDZ without CP ("raw") performs poorly due to a large error in the HF+CABS component, which is mitigated through CP. Again, without CP, ano-pVDZ+ outperforms the similarly-sized aug-pVDZ, but applying CP closes the gap.

In the absence of CP, VDZ-F12 clearly does best, but with CP places second after AVDZ. aVDZ-F12 has no clear advantage over VDZ-F12 "raw", but does somewhat improve with a better CABS. Its performance with partial and full CP, however, does offer an improvement over the underlying VDZ-F12.

Let us now turn to the triple-zeta options. sano-pVTZ is not very useful, while ano-pVTZ+ appears to be an improvement over AVTZ. The latter performs poorly in the absence of counterpoise corrections, but actually beats VTZ-F12 with counterpoise. aVTZ-F12, however, is now markedly superior over VTZ-F12, and indeed over the other options.

**TABLE 1.** RMS deviations of various basis sets and CABS for the S66 benchmark. MP2-F12/V{T,Q}Z-F12, full CP is used as the reference.

| Orbital Basis Set | $N_{bas}$[a] | CABS | raw HF+CABS | raw MP2-F12 | CP HF+CABS | CP MP2-F12 | half-CP HF+CABS | half-CP MP2-F12 |
|---|---|---|---|---|---|---|---|---|
| AVDZ | 384 | AVDZ/OptRI | 0.241 | 0.429 | 0.033 | 0.138 | 0.111 | 0.156 |
| sano-pVDZ | 300 | VDZ-F12/OptRI | 0.142 | 0.195 | 0.022 | 0.543 | 0.077 | 0.322 |
| ano-pVDZ+ | 396 | VDZ-F12/OptRI | 0.116 | 0.309 | 0.016 | 0.199 | 0.061 | 0.100 |
| VDZ-F12 | 468 | VDZ-F12/OptRI | 0.075 | 0.096 | 0.010 | 0.150 | 0.034 | 0.038 |
| aVDZ-F12 | 528 | VDZ-F12/OptRI | 0.069 | 0.097 | 0.010 | 0.085 | 0.032 | 0.029 |
| aVDZ-F12 | 528 | VDZ-F12/OptRI+ & diffuse | 0.049 | 0.078 | 0.004 | 0.090 | 0.024 | 0.028 |
| AVTZ | 828 | AVTZ/OptRI | 0.044 | 0.163 | 0.007 | 0.031 | 0.020 | 0.068 |
| sano-pVTZ | 600 | vanilla VTZ-F12/OptRI | 0.039 | 0.115 | 0.008 | 0.177 | 0.022 | 0.109 |
| ano-pVTZ+ | 840 | vanilla VTZ-F12/OptRI | 0.028 | 0.088 | 0.005 | 0.099 | 0.015 | 0.059 |
| VTZ-F12 | 852 | vanilla VTZ-F12/OptRI | 0.030 | 0.069 | 0.002 | 0.054 | 0.014 | 0.015 |
| aVTZ-F12 | 996 | vanilla VTZ-F12/OptRI | 0.019 | 0.045 | 0.002 | 0.019 | 0.009 | 0.014 |
| AVQZ | 1512 | AVQZ/OptRI | 0.019 | 0.073 | 0.002 | 0.015 | 0.009 | 0.030 |
| sano-pVQZ | 1080 | vanilla VQZ-F12/OptRI | 0.028 | 0.063 | 0.003 | 0.073 | 0.015 | 0.045 |
| ano-pVQZ+ | 1332 | vanilla VQZ-F12/OptRI | 0.024 | 0.056 | 0.003 | 0.050 | 0.013 | 0.031 |
| VQZ-F12 | 1452 | vanilla VQZ-F12/OptRI | 0.009 | 0.029 | 0.000 | 0.015 | 0.005 | 0.008 |
| aVQZ-F12 | 1704 | vanilla VQZ-F12/OptRI | 0.004 | 0.009 | 0.000 | 0.008 | 0.002 | 0.004 |
| aVQZ-F12 | 1704 | VQZ-F12/OptRI & diffuse | 0.003 | 0.007 | 0.000 | 0.008 | 0.001 | 0.004 |
| AV5Z | 2484 | AV5Z/OptRI | 0.002 | 0.019 | - | - | - | - |
| V5Z-F12 | 2316 | AVQZ/MP2FIT | 0.001 | 0.009 | 0.000 | 0.007 | 0.001 | 0.005 |

$N_{bas}$ is the number of basis functions for benzene dimer, by way of illustration

As for the quadruple-zeta options, aVQZ-F12 essentially yields V5Z-F12 quality at reduced cost. (It should be noted that we had to remove the diffuse *f* function on C for reasons of numerical stability.) With CP, the same is also true of VQZ-F12. sano-pVQZ is actually inferior in performance to aV**T**Z-F12, while ano-pVQZ+ may be preferable over AVQZ.

Finally, we find V5Z-F12 (and hence by extension, aVQZ-F12) to be markedly superior to AV5Z.

## Transferability of CABS

Using the VDZ-F12/OptRI in conjunction with ano-pVDZ+ clearly yields an unacceptable error, particularly raw. AVDZ/OptRI does noticeably better, its CABS error being a factor of five smaller than the basis set incompleteness error — which is actually numerically more precise than VDZ-F12/OptRI for VDZ-F12 itself.

For ano-pVTZ+, VTZ-F12/OptRI and AVTZ/OptRI are comparable in performance, and both cause smaller numerical errors than VTZ-F12/OptRI for VTZ-F12. For ano-pVQZ+, there is little to choose between AVQZ/OptRI and VTZ-F12/OptRI

**TABLE 2.** RMS deviations of various basis sets and CABS for the S66 benchmark. For each orbital basis set and CABS combination, the same orbital basis, combined with the aug-cc-pwCV5Z/MP2fit CABS, is used as the reference.

| Orbital Basis Set | CABS | $N_{CABS}$[b] | raw | | CP | |
|---|---|---|---|---|---|---|
| | | | HF+CABS | E2corr | HF+CABS | E2corr |
| ano-pVDZ+ | VDZ-F12/OptRI | 1056 | 0.043 | 0.131 | 0.003 | 0.038 |
| ano-pVDZ+ | AVDZ/OptRI | 1092 | 0.016 | 0.033 | 0.002 | 0.040 |
| ano-pVTZ+ | VTZ-F12/OptRI | 1392 | 0.006 | 0.003 | 0.001 | 0.005 |
| ano-pVTZ+ | AVTZ/OptRI | 1284 | 0.005 | 0.004 | 0.001 | 0.005 |
| ano-pVQZ+ | VQZ-F12/OptRI | 1632 | 0.008 | 0.004 | 0.001 | 0.004 |
| ano-pVQZ+ | AVQZ/OptRI | 1824 | 0.002 | 0.006 | 0.000 | 0.003 |
| VDZ-F12 | VDZ-F12/OptRI | 1056 | 0.029 | 0.010 | 0.012 | 0.016 |
| VDZ-F12 | VDZ-F12/OptRI+ | 1164 | 0.012 | 0.012 | 0.005 | 0.015 |
| VTZ-F12 | VTZ-F12/OptRI | 1392 | 0.017 | 0.002 | 0.003 | 0.002 |
| VTZ-F12 | VTZ-F12/OptRI+ | 1500 | 0.024 | 0.002 | 0.001 | 0.002 |
| VQZ-F12[a] | VQZ-F12/OptRI | 1632 | 0.006 | 0.001 | 0.000 | 0.000 |
| aVDZ-F12 | VDZ-F12/OptRI | 1056 | 0.029 | 0.011 | 0.012 | 0.014 |
| aVDZ-F12 | VDZ-F12/OptRI+ | 1164 | 0.010 | 0.013 | 0.006 | 0.013 |
| aVDZ-F12 | VDZ-F12/OptRI+ & diffuse | 1464 | 0.008 | 0.008 | 0.005 | 0.003 |
| aVTZ-F12 | VTZ-F12/OptRI | 1284 | 0.010 | 0.003 | 0.003 | 0.001 |
| aVTZ-F12 | VTZ-F12/OptRI+ | 1500 | 0.015 | 0.004 | 0.001 | 0.001 |
| aVTZ-F12 | VTZ-F12/OptRI+ & diffuse | 1692 | 0.006 | 0.004 | 0.001 | 0.000 |
| aVQZ-F12 | VQZ-F12/OptRI | 1632 | 0.002 | 0.001 | 0.000 | 0.001 |
| aVQZ-F12 | VQZ-F12/OptRI & diffuse | 1932 | 0.001 | 0.001 | 0.000 | 0.001 |

(a) number of CABS functions for benzene dimer, by way of illustration
(b) Using VQZ-F12/OptRI+ instead of VQZ-F12/OptRI produces virtually indistinguishable results.

Let us now turn to the recently proposed VnZ-F12/OptRI+ basis sets. With VDZ-F12, the added functions do improve the HF+CABS part for this basis set, but not really the MP2-F12 correlation energy. VTZ-F12/OptRI+ does enhance the HF+CABS part a bit for VTZ-F12 (only with CP), but no measurable effect of the additional functions on the correlation energy is seen. For VQZ-F12, the effect of the additional functions in OptRI+ is 0.001 kcal/mol or less for the S66 benchmark.

What about the aVnZ-F12 basis sets? For aVDZ-F12, the winning combination appears to be to add one layer of diffuse functions to the VDZ-F12/OptRI+ CABS basis set—the extra diffuse functions appear to be helpful for the correlation energy, while OptRI+ improves HF+CABS. (The exponents for the diffuse layer were obtained by "even-tempered"[25] extrapolation from the two outermost exponents.) The same holds true for aVTZ-F12 (though the difference is smaller here), while for aVQZ-F12, even the original VQZ-F12/OptRI is clearly adequate.

# CONCLUSIONS

From the S66 noncovalent interaction benchmark, we learn that the CABS is more transferable than generally assumed, at least between similar-sized basis sets: for instance, AVnZ/OptRI works acceptably well for ano-pVnZ+ and, by extension, for sano-pVnZ. For this benchmark, we see that OptRI+ presents a significant improvement over OptRI for VDZ-F12, but represents no significant improvement for VnZ-F12 (n=T,Q).

For the aVnZ-F12 (n=D,T) basis sets, an accurate CABS is obtained by adding a single diffuse layer to VnZ-F12/OptRI+, but the original VQZ-F12/OptRI is acceptable for aVQZ-F12.

Although aVnZ-F12 basis sets were initially developed with anionic systems in mind, they appear to generally speed up basis set convergence for noncovalent interactions, with aVQZ-F12 representing a de facto basis set limit at smaller cost than V5Z-F12.

# ACKNOWLEDGMENTS

This research was supported by the Israel Science Foundation (grant 1358/15), the Minerva Foundation, the Lise Meitner-Minerva Center for Computational Quantum Chemistry, and the Helen and Martin Kimmel Center for Molecular Design (Weizmann Institute of Science).